\documentclass[pdflatex,sn-mathphys-num]{sn-jnl}

\usepackage{graphicx}%
\usepackage{multirow}%
\usepackage{amsmath,amssymb,amsfonts}%
\usepackage{amsthm}%
\usepackage{mathrsfs}%
\usepackage[title]{appendix}%
\usepackage{xcolor}%
\usepackage{textcomp}%
\usepackage{manyfoot}%
\usepackage{booktabs}%
\usepackage{algorithm}%
\usepackage{algorithmicx}%
\usepackage{algpseudocode}%
\usepackage{listings}%
\usepackage{mathtools}

\usepackage{color,xcolor,ucs}
\usepackage{mathtools}   \usepackage{tikz} 
\usepackage{ amssymb }
\usepackage{extarrows} 
\usepackage{pgf,tikz}
\usepackage{float}
\usetikzlibrary{positioning}
\usetikzlibrary{shapes.geometric}
\usetikzlibrary{shapes.misc}
\usetikzlibrary{arrows}
\usepackage{caption}
\usepackage{mathrsfs}
\usetikzlibrary{arrows,shapes,automata,backgrounds,petri,positioning}
\usetikzlibrary{decorations.pathmorphing}
\usetikzlibrary{decorations.shapes}
\usetikzlibrary{decorations.text}
\usetikzlibrary{decorations.fractals}
\usetikzlibrary{decorations.footprints}
\usetikzlibrary{shadows}
\usetikzlibrary{calc}
\usetikzlibrary{spy}
\usepackage{amsmath}
\usepackage{array}
\usepackage{ amssymb }
\usepackage{braket}
\usepackage{qcircuit}
\usepackage{soul}
\usepackage{braket} 
\usepackage{relsize}

\usepackage{amsmath}
\usepackage{ amssymb }
\usepackage{braket}
\usepackage{qcircuit}
\usepackage{soul}
\usepackage{braket}

\theoremstyle{thmstyleone}%
\theoremstyle{thmstyletwo}%

\theoremstyle{thmstylethree}%

\begin{document}

\title[Article Title]{Eve's forgery probability from her false acceptance probability:  interactive authentication, Holevo information and the min-entropy}


\author*{\fnm{Pete} \sur{Rigas}}\email{pbr43@cornell.edu}

\affil*{Newport Beach, CA 92625}

\abstract{We obtain estimates for Eve's forgery probability, namely the probability that she is able to forge a message which Alice or Bob mistakenly accept over a noisy Quantum channel for generating a shared Quantum secret key. This probability is related to Eve's success probability obtained in a previous work due to Renner and Wolf, which was obtained from assumptions on the min-entropy for characterizing asymmetric security. To demonstrate that protocols over noisy Quantum channels are dependent upon a single, unified security threshold in comparison to multiple security parameters in the Renner-Wolf interactive authentication protocol framework we upper bound Eve's forgery probability with a Holevo-type quantity that can be made negligibly small. By leveraging estimates for Eve's false acceptance probability that have previously been obtained by the author, we obtain the desired security threshold by bounding the false acceptance probability with a suitably chosen two-universal function which serves as a counterpart to two-universal hashing functions that have previously been exained for cryptographic protocols in Quantum key distribution. As a result the protocol is not only $\epsilon$-secure, for some $\epsilon>0$, but also composable against forgery and key leakage. \textit{{Keywords}: Quantum games, non-locality, Quantum computation} \footnote{\textbf{MSC Class}: 81P02; 81Q02}}

\maketitle

\section{Introduction}

\subsection{Overview}

Unconditional security in Quantum Information theory claims, irrespective of the computational resources of an eavesdropper who is usually taken to be Eve, that no aspects of the information transmitted between Adam and Eve can be intercepted. Over Classical and Quantum channels alike, previous works of the author have thoroughly characterized possible ways in which Eve can compromise the security of information that is transmitted between Alice and Bob, either for the purposes of: (1) transmitting secret Quantum keys, [21], composability, [22], parallel repetition, in which classical bits of information are transmitted through a suitable encoding by Alice or Bob repeatedly over a Quantum channel, [20], error correction, [19], and approximately, or exactly, optimal strategies for games [18]. Albeit the fact that previous research efforts within the Quantum Information community have established computational tasks with prospective Quantum advantage ranging from, aerospace applications, [2], winning probabilities of Quantum games, [3], learning algorithms, [4,5,6], approximate optimization, [7], and entanglement [8], additional possibilities remain for characterizing \textit{asymmetry} in security, as identified in [16]. In the context of security, \textit{asymmetry}, as described in [16], captures how Alice or Bob may have more security for the purposes of cryptography than the other participant in comparison to previous works which typically assume that Alice and Bob have \textit{symmetric} security. Within this Quantum framework, enforcing \textit{asymmetry}, rather than \textit{symmetry}, for cryptographic protocols is leveraged for concluding that primitives can be much weaker than initially thought for unconditional security.

To further build upon asymmetric security described in [16], we not only draw upon the authors' arguments relating to authentication, information reconciliation, and privacy amplification, but also formulate how such protocols relate to previous arguments of Quantum Key Distribution (QKD) independent protocols [10]. Generally speaking, determining whether aspects of QKD independent protocols can be incorporated into information processing tasks with QKD dependent protocols allows for one to: determine whether classical, or quantum, bits of information transmitted by Alice or Bob are unconditionally secure; characterize possible sources of asymmetric in security, specifically through possible sources of cryptographic security favoring Alice's security over Bob's security, or vice versa; potential connections with the computational complexity of optimization, [1], variational quantum algorithms of stochastic PDEs, [9,12], QKD paramter estimation, [11], and several related topics [13,14,15]. Moreover, investigating these topics of interest could lead to further generalizations of security described in [11], and further characterized in [21], in which the author quantified how the action of isometries impacts the security threshold for hashing algorithms that are used for extracting secure Quantum keys with high probability and hence with high security.

To this end, we quantify how previous expressions for Eve's false acceptance probability that have been obtained in [22] can be used to study Eve's forgery probability. As a general matter of fact, the false acceptance and forgery probabilities of an adversary are always of great importance to examine in cryptographic protocols; besides having a noticeable impact on the security thresholds of both QKD independent, and dependent, protocols, the false acceptance and forgery probabilities are intimately related to how Eve can compromise the security of many cryptographic protocols. In previous settings, while Alice and Bob can significantly increase Eve's false acceptance probability through cascading over the forward and backward conceptualizations of public broadcast channels, in turn increasing the security of their protocol without a Quantum secret key, additional possibilities remain. For one, it could be of interest to examine how Eve could mount attacks against cryptographic protocols that are dependent upon Authentication, Information Reconciliation, and Privacy Amplification, protocols as provided in [16]. Together these protocols have been widely used throughout the field of Quantum Information theory for providing quantitative security estimates for QKD protocols. One such security estimate, along with its connections with Alice's and Bob's isometries, have been derived in [21].

Initially, even if there does not appear to be a connection between Authentication, Information Reconciliation, and Privacy Amplification protocols and cascading over the public broadcast channel, being able to obtain bounds on the forgery probability from the false acceptance probability is of great interest to formalize, as it would: (1) illustrate how the Holevo information, along with channel degradation mechanisms, can be exploited by Eve to turn her false acceptance \textit{disadvantage} to a forgery \textit{advantage}; (2) achieving optimal performance of cryptographic protocols, in the sense of Alice and Bob being able to agree upon a shared Quantum secret key if the hashing protocol is not aborted; (3) establishing how unconditional security can be achieved under a wide variety of assumptions, ranging from QKD cryptographic protocols with the use of a hashing function with \textit{symmetric}, unconditional security, or otherwise weaker classes of primitives for \textit{asymmetric}, unconditional security.

With regards to \textit{symmetric}, unconditional, and \textit{asymmetric}, unconditional security, determining the conditions under which secure cryptographic protocols can be obtained is of great interest. Although unconditional security has been shown to hold in a wide variety of cryptographic circumstances, whether such strong theoretical guarantees hold if Alice and Bob do not have equal access to authenticity and confidentiality resources (related to the so called "channel calculus" described in [16]) would provide new implications on the Holevo information and related objects from [22]. Altogether, reformulating the conditions that have been provided in [16] with the smooth relative min entropy with the Holevo information can be leveraged for the purposes of: intepreting the initial correlated weak randomness between Alice and Bob in the Renato-Wolf framework as a specialization of the noisy Quantum channel analyzed in [19]; augmenting min-entropy measures with the Mutual Information entropy, and Holevo information analyzed in [22]; incorporating the false acceptance probability lower bound in [22] for bounding the probabilty that Alice or Bob mistakenly accept a forged message from Eve; obtaining the bound for Eve's forgery probability from bounds on Eve's success probability in the Renato-Wolf interactive authentication protocol; establishing connections between combinatorial polynomial challenge responses through min-entropy analyses to information-theoretic authentication bounds through channel capacity results and the Holevo information.

At the center of this approach are several additional observations related to the post-processing the images of Quantum states under Completely Positive Trace Preserving (CPTP) maps. Such maps were first put to significant use in [22] for not only demonstrating, through data-processing inequalities, that the entropy between two states is lower bounded by the entropy of the images of the states under CPTP maps, but also for increasing Eve's uncertainty through cascading of the public broadcast channel. Despite the fact that we are not directly making use of the public broadcast channel in this setting, it nevertheless serves as an informative construction for tightening Eve's uncertainty through the forgery probability. Specifically within the context of the Renato-Wolf interactive authentication protocol a Holevo-type quantity can be used to upper bound the probability that Eve is able to sustain some form of cryptographic advantage through forgery hence translating to a negligible false acceptance rate. It should not come at a surprise that a Holevo-type quantity, much like the Holevo-type sums manipulated in [22] for quantifying the degradation of a Quantum channel, again come into play except within the context of the interactive authentication protocol. 

Given this overview, in the next section we provide a general description of the objects associated with the role of the Holevo information in the interactive authentication protocol. In particular, we draw the attention of the reader to Holevo-type sums that were used to provide an up to constants lower bound for Eve's false acceptance probability, in addition to the the degradation mechanisms of Quantum channels through letter reduction. However, beyond these previous uses of the Holevo information and CPTP maps, we highlight novel prospects for obtaining estimates on Eve's forgery probability. The forgery probability, in comparison to the false acceptance probability, from Eve's perspective is intricately related to the information-theoretic primitive which Alice and Bob make use of when initiating the cryptographic protocol. The role of such primitives introduced within the Renato-Wolf framework has not been analyzed for QKD independent protocols in [22]; it is therefore of great interest to make use of the unconditional, and composable, aspects of security with such information-theoretic primitives.

\subsection{This paper's contributions}

In this paper we further elucidate upon the notion of authentication as an information-theoretic primitive. Crucial to the forthcoming approach include: (1) an overview of the Holevo information and Holevo-type summations used with post-processing of CPTP maps; (2) quantifying Eve's prospective advantage to Alice and Bob in the cryptographic protocol after interaction; (3) broadening the authentication protocol introduced in the Renato-Wolf framework for QKD independent protocols; (4) composing a channel with guaranteed false acceptance properties with key extraction in the Renato-Wolf framework, therefore permitting for a full key agreement between Alice and Bob over an authenticated channel with composable security. In the next section below we begin to introduce objects related to this approach.

\subsection{Paper organization}

We present several Cryptographic-theoretic objects in the next section before stating the main results in Section \textit{2.4}. Specifically: in Section \textit{2.1} we demonstrate how game-theoretic objects previously analyzed by the author in [17,18,19,20] can be leveraged towards obtaining estimates on Eve's false acceptance probability over the forwards conceptual public broadcast channel; in Section \textit{2.2} we demonstrate how objects relating to the Parity check matrices can be manipulated within a QKD hashing protocol and its corresponding security threshold obtained in [21], as an extension of a security threshold previously identified in [11]; in Section \textit{2.3} we demonstrate how the min entropy, along with classical shared randomness introduced in [16], can be employed with authentication, information reconciliation and privacy amplification protocols for characterizing asymmetry of security. In light of asymmetric security identified in [16], in Section \textit{2.4} we combine properties of the Holevo information used for lower bounding Eve's false acceptance probability with properties of the min entropy in the Renner-Wolf framework, hence obtaining a bound for Eve's forgery probability in terms of her false acceptance probability. In Section \textit{3} we present the arguments for the two main results, not only demonstrating how replacing the min entropy with the Holevo information allows for an estimate on Eve's forgery probability, but also how the interactive authentication protocol in the Renner-Wolf framework can be analyzed with a single, unified security threshold $\epsilon$.

\section{Cryptographic-theoretic objects}

\noindent We define several cryptographically oriented objects before stating the main results.

\subsection{Composable, unconditional security independent of a Quantum secret key}

\noindent We introduce a few objects from [22].

\bigskip

\noindent \textbf{Definition} \textit{1} (\textit{the Frobenius norm}). Denote the Frobenius norm with,

\begin{align*}
  \big|\big| A \big|\big|_F \equiv \sqrt{\overset{m}{\underset{i=1}{\sum}} \overset{n}{\underset{j=1}{\sum}} \big| a_{ij} \big|^2 } = \sqrt{\mathrm{Tr} \big[ A^{\dagger} A \big] }  \text{, } 
\end{align*}

\noindent of an $m \times n$ matrix $A$ with entries $a_{ij}$, where $\dagger$ denotes the complex conjugate transpose.

\bigskip

\noindent \textbf{Definition} \textit{2} (\textit{linear bijection}). Fix $d_A \neq d_B >0 $ which respectively correspond to the dimensions of Alice's space and Bob's space, respectively. There exists a \textit{linear bijection} $\mathcal{L}$ between the tensor product space, $\textbf{C}^{d_A} \otimes \textbf{C}^{d_B}$, and the space of $d_A \times d_B$ matrices with complex entries, $\mathrm{Mat}_{d_A , d_B} \big( \textbf{C} \big)$, satisfying (\textbf{Lemma} \textit{1}, {\color{blue}[38]}),

\begin{itemize}
\item[$\bullet$] \underline{\textit{Image of the tensor product of two quantum states under} $\mathcal{L}$}: $\forall \ket{u} \in \textbf{C}^{d_A}, \ket{w} \in \textbf{C}^{d_B}, \exists \ket{u^{*}} \in \textbf{C}^{d_B} : \mathcal{L} \big( \ket{u} \otimes \ket{w} \big) = \ket{u} \bra{u^{*}}  \text{, }$ 
\item[$\bullet$] \underline{\textit{Product of a matrix with the image of a quantum state under} $\mathcal{L}$}: $\forall \ket{u} \in \textbf{C}^{d_A}, \exists A \in \mathrm{Mat}_{d_A} \big( \textbf{C} \big) : A \mathcal{L} \big( \ket{u} \big) = \mathcal{L} \big( A \otimes I \ket{u} \big)\text{, }$
\item[$\bullet$] \underline{\textit{Product of the image of a quantum state under $\mathcal{L}$ with the transpose of a matrix}}:  $\forall \ket{w} \in \textbf{C}^{d_B}, \exists B \in \mathrm{Mat}_{d_B} \big( \textbf{C} \big) : \mathcal{L} \big( \ket{w} \big) B^T = \mathcal{L} \big( I \otimes B \ket{w} \big)  \text{, }$
\item[$\bullet$] \underline{\textit{Frobenius norm equality}}: $\forall \ket{w} \in \textbf{C}^{d_B} : \big|\big| \mathcal{L} \big(   \ket{w}     \big) 
 \big|\big|_F = \ket{w}  \text{. } $
\end{itemize}

\noindent where the basis of $\textbf{C}^{d_A} \otimes \textbf{C}^{d_B}$ is of the form $\ket{i} \otimes \ket{j}$,  and the basis for $\mathrm{Mat}_{d_A, d_B} \big( \textbf{C} \big)$ is of the form $\ket{i}\bra{j}$, for $1 \leq i \leq d_A$ and $1 \leq j \leq d_B$.

\bigskip

\noindent \textbf{Definition} \textit{3} (\textit{the answer sets that players use in two-player game-theoretic settings}). Denote,

\begin{align*}
 A_S \equiv \underset{s \in S}{\bigcup} A_s \equiv \big\{   s  \in S :  A_s \in \big\{ - 1 , + 1 \big\}    \big\}     \text{, } \\    B_T \equiv \underset{t \in T}{\bigcup} B_t \equiv \big\{ t \in T  :  B_t \in \big\{ -1 , + 1 \big\}  \big\}    \text{, }
\end{align*}

\noindent corresponding to the set of sets of answers that Alice and Bob, respectively, can provide in respond to a question drawn uniformly at random from the referree's probability distribution $\pi$.

\bigskip

\noindent \textbf{Definition} \textit{4} (\textit{the scoring predicate for the referee}). From a product probability distribution $\pi$ over $S \times T$, referee presiding over a two-player Quantum game examines the responses of Alice and Bob depending upon the entangled state that they share, in which, after sampling a pair $\big( S , T \big) \sim \pi$, and sending one question $s$ to Alice and another question $t$ to Bob,

\begin{align*}
    V \big(  s , t \big)   ab  \equiv 1 \Longleftrightarrow  \text{ Alice and Bob win,}    \\    V \big(  s , t \big)   ab \equiv -1 \Longleftrightarrow  \text{ Alice and Bob lose,}      
\end{align*}

\noindent in which, depending upon whether $V \big( s ,t \big) \equiv 1$, or $V \big( s , t \big) \equiv -1$, Alice and Bob must either give the same answers, and opposing answers, to win, respectively.

\bigskip

\noindent To further illustrate how game-theoretic objects relate to the probability of Alice and Bob winning a Quantum game, which can be directly applied to the CHSH/XOR, [10,11], or even to the multiplayer counterpart of the XOR and $\mathrm{XOR}^{*}$ games, [17,18,19,20], we formulate how Alice and Bob can make use of suitably chosen decoding and encoding schemes for transmitting Classical information over a Quantum channel.

\bigskip

\noindent \textbf{Definition} \textit{5} (\textit{cryptographic objects over Quantum channels}). Denote,

{\small \[  \left\{\!\begin{array}{ll@{}>{{}}l} 
      \mathcal{R} \neq \mathcal{S} \equiv \text{Two resources which take inputs from Alice, Bob and Eve}    \text{, } \\ \\               \mathcal{R} \big| \big| \mathcal{S} \equiv \text{A resource of } \mathcal{R}  \text{ and } \mathcal{S}  \text{ simultaneously}           \text{, } \\ \\   d \big( \cdot , \cdot \big) \equiv \text{A metric between two resources}     \text{, } \\ \\  \mathcal{N}^n_{p,q} \equiv \text{Alice and Bob's resource for sharing } n\text{-bit authenticated messages over}   \\ \text{the Quantum channel with Bernoulli}-p\text{ random variables}    \text{, }  \\ \\  \mathcal{A}^{rn} \equiv    \text{Alice and Bob's resource for sharing } n\text{-bit authenticated message over the} \\ \text{Quantum channel}       \text{. }
\end{array}\right. \] } 


\bigskip

\noindent \textbf{Definition} \textit{6} (\textit{various probabilities related to error correction and false acceptance over a Quantum channel}). Denote the probabilities,

{\small \[  \left\{\!\begin{array}{ll@{}>{{}}l} 
 \underline{P_{\mathrm{EC}, A }} \equiv         \underset{A }{\underset{\mathrm{ec} \in \mathrm{EC}}{\mathrm{sup}}}    p_{\mathrm{ec}} \equiv          \mathrm{sup} \big\{ \text{success probability of a player using an error cor-} \\ \text{ recting code over Alice's independent Quantum channel} \big\}               \text{, }  \\ \\  \underline{P_{\mathrm{FA}, A }} \equiv           \underset{A }{\underset{\mathrm{fa} \in \mathrm{FA}}{\mathrm{inf}}}    p_{\mathrm{fa}} \equiv           \mathrm{inf} \big\{ \text{failure probability of a player accepting a message that} \\    \text{  should have not been accepted over Alice's independent Quantum channel} \big\}                    \text{, } \\ \\ \underline{P_{\mathrm{EC}, B }} \equiv         \underset{B }{\underset{\mathrm{ec} \in \mathrm{EC}}{\mathrm{sup}}}    p_{\mathrm{ec}} \equiv          \mathrm{sup} \big\{ \text{success probability of a player using an error correcting} \\ \text{ code over Bob's independent Quantum channel} \big\}               \text{, }  \\ \\   \underline{P_{\mathrm{FA}, B }} \equiv           \underset{B}{\underset{\mathrm{fa} \in \mathrm{FA}}{\mathrm{inf}}}    p_{\mathrm{fa}} \equiv           \mathrm{inf} \big\{ \text{failure probability of a player accepting a mes-} \\     \text{ sage that should have not been accepted over Bob's independent} \\  \text{ Quantum channel} \big\}                    \text{, }  \\ \\   \underline{P_{\mathrm{EC}, E}} \equiv         \underset{E}{\underset{\mathrm{ec} \in \mathrm{EC}}{\mathrm{sup}}}    p_{\mathrm{ec}} \equiv          \mathrm{sup} \big\{ \text{success probability of a player using an error cor} 
 \\   \text{ -recting code over Eve's independent Quantum channel} \big\}               \text{, }  \\ \\  \underline{P_{\mathrm{FA}, E}} \equiv           \underset{E}{\underset{\mathrm{fa} \in \mathrm{FA}}{\mathrm{inf}}}    p_{\mathrm{fa}} \equiv           \mathrm{inf} \big\{ \text{failure probability of a player accepting a message that} \\ \text{ should  have not been accepted over Eve's independent Quantum channel} \big\}                    \text{, }
\end{array}\right. \] } 


\noindent corresponding to the occurrence of error correction and false acceptance, of Alice and Bob over a Quantum channel, respectively.

\bigskip

\noindent \textbf{Theorem} (\textbf{Theorem} \textit{1}, [22]). Denote, $N \neq M > 0$, Eve's Holevo information, $\chi_E$ and the binary entropy function $H_b \big( p \big) : = p \mathrm{log} \big( p \big) + \big( 1 - p \big) \mathrm{log} \big( 1 - p \big)$. The following properties of $p_{\mathrm{FA},\mathrm{E}}$, Eve's false acceptance probability over the Quantum channel, hold:

\begin{itemize}

 \item [$\bullet$] \textit{Minimization of probabilities via the Fano inequality}. One has that,

\begin{align*}
  p_{\mathrm{FA},\mathrm{E}} \equiv \underset{\chi_E > 0}{\underset{M > 0}{\bigcup}}  \mathrm{inf} \big\{ p :   \big\{ p > 0 \big\} , \big\{    \mathrm{log} M - \chi_E \leq  H_b \big(  p \big)   + p \mathrm{log}  \big[ M - 1 \big]    \big\}       \big\}   .
\end{align*}

\item[$\bullet$] \textit{Up to constant lowers bounds for Eve's error probability}. One has that,

\begin{align*}
  P_{\mathrm{FA},\mathrm{E}} \gtrsim  \mathrm{log} \bigg[ \frac{1}{N} \bigg] \bigg[ \mathrm{log} M - \chi_E - 1 \bigg]  ,
\end{align*}

\noindent for $N \neq M$, each of which can be taken to be sufficiently large.

\bigskip

\item[$\bullet$] \textit{Sharpening the up to constants lower bound for Eve's error probability to an inequality via Helstrom}. One has that,

\begin{align*}
    P_{\mathrm{FA}, \mathrm{E}} \geq 1 - \frac{1 + \epsilon^{\prime\prime} \big( M - 1 \big)}{M}      , 
\end{align*}

\noindent for $\epsilon^{\prime\prime}$ taken to be sufficiently small.

\end{itemize}

\subsection{Composable security dependent on a Quantum secret key}

\noindent We introduce a few objects from [21].

\bigskip

\noindent \textbf{Definition} \textit{7} (\textit{parity check matrices from the marginals of probability distributions supported over random matrices}). The parity check matrices, as the marginal distributions,

\begin{align*}
     \textbf{P}_L \big[ \cdot \big] \bigg|_{\textit{first column}} \equiv \mathcal{P}_1   , \\ \\ \textbf{P}_{(L^{-1} )^{\mathrm{T}}} \big[ \cdot \big] \bigg|_{\textit{second column}} \equiv \mathcal{P}_2  , 
\end{align*}

\noindent with support given by the random matrices $L,\big( L^{-1} \big)^{\mathrm{T}}$ over the field with two elements are given as $\mathcal{P}_1$ and $\mathcal{P}_2$, respectively.

\bigskip

\noindent \textbf{Definition} \textit{8} (\textit{Pauli operations}). Denote,

\begin{align*}
   \sigma_1 =    \begin{bmatrix}
   0 & 1   \\ 1 & 0 
    \end{bmatrix}  , \\ \\   \sigma_3 = \begin{bmatrix}
    1 & 0  \\ 0 & -1 
    \end{bmatrix} , 
\end{align*}

\noindent corresponding to the Pauli operations $\sigma_1$ and $\sigma_3$, respectively.

\bigskip

\noindent We now define the QKD hashing protocol. This is closely related to the role of the two-universal functions introduced in \textbf{Definition} \textit{15}, which in [16] serve the purpose of identifying asymmetry in composable security.

\bigskip

\noindent \textbf{Definition} \textit{9} (\textit{hashing protocols for Quantum key distribution}). Denote,

{\small \begin{align*}
  x_A \equiv \textit{Alice's measured state that she obtains in the $\ket{+}$, $\ket{-}$ computational basis from Eve's state}  , \\ \\ x_B \equiv \textit{Bob's measured state that she obtains in the $\ket{+}$, $\ket{-}$ computational basis from Eve's state} , \\ \\ z_A \equiv    \textit{Alice's measured state that is initially distributed to her at the beginning of the hashing pro-} \\ \textit{tocol by Eve}        , \\ \\ z_B \equiv    \textit{Bob's measured state that is initially distributed to her at the beginning of the hashing pro-} \\ \textit{tocol by Eve}         . 
\end{align*} }

\noindent For a QKD protocol $\pi^{\prime} \equiv \pi^{\prime} \big( n , k ,r \big)$, where $n$ denotes the total number of qubits in the state which Eve initially distributes to Alice and Bob, $k$ is the number of ancilla qubits in the state which Eve initially distributes to Alice and Bob, and $r$ is half of the number of phase flip, and bit flip, errors, the hashing protocol,

{\small \[ \left\{\!\begin{array}{ll@{}l} 
\textbf{Initialization}, \textit{(0)}: \textit{Alice and Bob initialize a system, } AB, \textit{ that they share between themselves,} \\ \textit{and  which they can compute the trace of with the operator } \mathrm{Tr}_{AB} \big[ \cdot \big], \textit{from n qubit states that} \\   \textit{ they receive from Eve}.  \\ \\  \textbf{Random matrix state purification}, \textit{(1)}: \textit{Alice and Bob pick a random matrix}, L, \textit{which they purify} \\ \textit{to } \ket{\mathcal{L}}_{\textbf{L} \textbf{L}^{\prime}} \textit{ and which they can compute the trace of with the operator }  \mathrm{Tr}_{\textbf{L}^{\prime}} \big[ \cdot \big].  \\ \\ \textbf{Isometry}, \textit{(2)}: \textit{Alice and Bob apply the isometry } L_1 \textit{ to } z_A, \textit{ and to } z_B, \textit{respectively, which they can } \\ \textit{compute the trace of with the operator } \mathrm{Tr}_{U^{\prime}_A U^{\prime}_B} \big[ \cdot \big]  .  \\ \\ \textbf{Transpose of random matrix inverse}, \textit{(3)}:     \textit{Alice and Bob apply the second, and third, columns of } 
\\  \big( L^{-1} \big)^{\mathrm{T}} \textit{ to the results obtained from the previous step above, which they can compute the trace of} 
\\  \textit{with the operator } \mathrm{Tr}_{V^{\prime}_A V^{\prime}_B} \big[ \cdot \big]     . \\ \\ \textbf{Parity check matrices computations}, \textit{(4)}:        \textit{Alice and Bob compute functions } g_1 \textit{ and } g_2 \textit{ of } \\  \textit{ the parity check matrices, which they can compute the trace of with the operator } \mathrm{Tr}_{S^{\prime} T^{\prime}} \big[ \cdot \big] , \\ \\ \textbf{Secret key output}, \textit{(5)}: \textit{If the two-universal hashing protocol is not aborted, Alice and Bob output} \end{array}\right. \]
\[  \left\{\!\begin{array}{ll@{}>{{}}l}    \textit{ the secret key, which she first outputs as } w_B, \textit{and after which Bob accepts } w_B \\ + \big[ \textit{third column of } \big( L^{-1} \big)^{\mathrm{T}} \big] t \textit{ as a copy of, which they can compute the trace of} \\ \textit{ with the operator } \mathrm{Tr}_{W^{\prime}_A W^{\prime}_B} \big[ \cdot \big] .     
\end{array}\right. 
\]  }

\noindent associated with $\pi^{\prime}$ is dependent upon the computation of the states,

\begin{align*}
 \ket{\mathcal{P}_1  z_A,  \mathcal{P}_1 z_A , \mathcal{P}_1  z_B,  \mathcal{P}_1 z_B ,  g_1 \big( \mathcal{P}_1 , \mathcal{P}_1 \big( z_A + z_B \big)  \big)    , \mathcal{P}_1 z_B     }_{U_A U^{\prime}_A U_B U^{\prime}_B S U^{\prime\prime}_B}      , \\ \\    \ket{\mathcal{P}_1  z_A,  \mathcal{P}_1 z_A , \mathcal{P}_1  z_B,  \mathcal{P}_1 z_B , \mathcal{P}_1 z_B   ,  g_1 \big( \mathcal{P}_1 , \mathcal{P}_1 \big( z_A + z_B \big) \big)              }_{U_A U^{\prime}_A U_B U^{\prime}_B  U_B U^{\prime\prime}_B  S }       , \\ \\    \ket{\mathcal{P}_1  z_A,  \mathcal{P}_1 z_A , \mathcal{P}_1  z_B,  \mathcal{P}_1 z_B ,  \mathcal{P}_1  z_A     , \mathcal{P}_1 z_B     }_{U_A U^{\prime}_A U_B U^{\prime}_B U^{\prime\prime}_A U^{\prime\prime}_B}          ,  \\ \\   \ket{\mathcal{P}_1  z_A,  \mathcal{P}_1 z_A , \mathcal{P}_1  z_B,  \mathcal{P}_1 z_B , \mathcal{P}_1 z_B   ,  \mathcal{P}_1  z_A }_{U_A U^{\prime}_A U_B U^{\prime}_B U^{\prime\prime}_B U^{\prime\prime}_A  }        , \\ \\ \ket{\mathcal{P}_2  x_A,  \mathcal{P}_2 x_A , \mathcal{P}_2  x_B,  \mathcal{P}_2 x_B ,  g_2 \big( \mathcal{P}_1 , \mathcal{P}_2 \big( x_A + x_B \big) \big)  ,  \mathcal{P}_2  x_A  }_{U_A U^{\prime}_A U_B U^{\prime}_B T U^{\prime\prime}_A }       , \\ \\   \ket{\mathcal{P}_2  x_A,  \mathcal{P}_2 x_A , \mathcal{P}_2  x_B,  \mathcal{P}_2 x_B , \mathcal{P}_2 x_B   ,  g_2 \big( \mathcal{P}_1 , \mathcal{P}_2 \big( x_A + x_B \big) \big)         }_{U_A U^{\prime}_A U_B     U^{\prime}_B U^{\prime\prime}_A T     }    , \\ \\  \ket{\mathcal{P}_2  x_A,  \mathcal{P}_2 x_A , \mathcal{P}_2  x_B,  \mathcal{P}_2 x_B , \mathcal{P}_2  x_A  ,  \mathcal{P}_2  x_B  }_{U_A U^{\prime}_A U_B U^{\prime}_B U^{\prime\prime}_A U^{\prime\prime}_B }  , \\ \\   \ket{\mathcal{P}_2  x_A,  \mathcal{P}_2 x_A , \mathcal{P}_2  x_B,  \mathcal{P}_2 x_B , \mathcal{P}_2 x_B   ,  \mathcal{P}_2 x_A }_{U_A U^{\prime}_A U_B U^{\prime}_B U^{\prime\prime}_B U^{\prime\prime\prime}_B U^{\prime\prime}_A }       , 
\end{align*} 

\noindent where,

\begin{align*}
  g_1 \big[ \cdot \big] : = \textit{Function of $\mathcal{P}_1$}  , \\ \\ g_2 \big[ \cdot \big] : = \textit{Function of $\mathcal{P}_2$} . 
\end{align*}

\bigskip

\noindent \textbf{Definition} \textit{10} (\textit{POVMs corresponding to the two parity check matrices in the QKD hashing protocol}). Given a state $\ket{\psi} \equiv   \ket{\psi_{\alpha \beta}} \equiv \big[ \textbf{I} \otimes   \sigma^{\alpha^{\mathrm{T}}}_1 \sigma^{\beta^{\mathrm{T}}}_3       \big] \ket{\psi}$ supported over the field with two elements, in addition to a function $F$ of the set of anticipated errors in the QKD hashing protocol from bit flips and phase flips the isometries,

{\small \begin{align*}
    \mathscr{U} \equiv     \underset{\alpha \neq \beta^{\prime} \in \textbf{F}^n_2}{\sum}      \ket{\psi_{\alpha \beta^{\prime} }} \bra{\psi_{\alpha \beta^{\prime} }}  \otimes    \frac{1}{\sqrt{2}} \bigg[  \ket{F \big( \alpha \big) , F \big( \alpha + \beta^{\prime} \big)}_{S(S^{\prime} + T^{\prime})} +    \ket{F \big( \beta  \big) , F \big( \alpha + \beta^{\prime} \big)}_{T(S^{\prime} + T^{\prime})}    \bigg]     , \\ \\       \mathscr{V} \equiv        \underset{\alpha^{\prime} \neq \beta^{\prime\prime} \in \textbf{F}^n_2}{\sum}  \ket{\psi_{\alpha^{\prime}  \beta^{\prime\prime} }} \bra{\psi_{\alpha^{\prime} \beta^{\prime\prime} }} \otimes   \frac{1}{\sqrt{2}} \bigg[   \ket{F ( \alpha^{\prime} + \beta^{\prime\prime} ) , F \big( \alpha^{\prime}  \big) }_{(S+T) S^{\prime}} +   \ket{F ( \alpha^{\prime} + \beta^{\prime\prime} ) , F \big( \beta^{\prime\prime} \big) }_{(S+T) T^{\prime}}                \bigg]  ,  
\end{align*} }

\noindent the Positive Operator Valued Measurements (POVMs) corresponding to $\mathcal{P}_1$ and $\mathcal{P}_2$ above are denoted with,

{\small \begin{align*}
     \textbf{P}_L \big[  \textit{There exists a CSS code such that the first parity check matrix, from the finite dimensional} \\ \textit{representation of L, which can be used for error correction}       \big]    \equiv  \mathscr{P} \mathscr{O} \mathcal{V}  \mathscr{M}_{\mathscr{U}}      , 
\end{align*} } 

\noindent where,

\begin{align*}
 \underset{\mathscr{U}}{\sum } \mathscr{P} \mathscr{O} \mathcal{V}  \mathscr{M}_{\mathscr{U}} = \textbf{I}   , 
\end{align*}

\noindent and,

{\small \begin{align*}
       \textbf{P}_{ ( L^{-1} )^{\mathrm{T}}} \big[  \textit{There exists a CSS code such that the second parity check matrix, from the finite dimensional} \\ \textit{representation of $\big( L^{-1} \big)^{\mathrm{T}}$, which can be used for error correction}          \big]  \equiv          \mathscr{P} \mathscr{O} \mathcal{V}  \mathscr{M}_{\mathscr{V}}         , 
\end{align*} }

\noindent where,

\begin{align*}
   \underset{\mathscr{V}}{\sum } \mathscr{P} \mathscr{O} \mathcal{V}  \mathscr{M}_{\mathscr{V}} = \textbf{I}  . 
\end{align*}

\bigskip

\noindent \textbf{Theorem} (\textbf{Theorem} \textit{2}, [21]). Denote the hashing protocol with $\pi^{\prime} \big( n,k,r\big)$, along with the some $C>0$. For the number of bit flip, and phase flip, errors $2r$, and total number of qubits $n$ that are encoded in states that Alice and Bob initially input to the QKD hashing protocol, the protocol is,

\begin{align*}
  2^{-\frac{k}{2} + n \frac{h}{2} ( \frac{r}{n} ) + \frac{5}{2} ( 5 - \frac{3}{2} ) + \mathrm{log}_2 \sqrt{C}}  ,
\end{align*}

\noindent secure.

\subsection{min entropy and classical shared randomness dependent on a Quantum secret key}

\noindent We introduce a few objects from [16].

\bigskip

\noindent \textbf{Definition} \textit{11} (\textit{weak information-theoretic primitives}, \textit{1.1} [16]). Denote an information-theoretic primitive $I$ as the information shared between Alice and Bob before a QKD protocol is initiated (often taken as some portion of qubits in a secret key that is much shorter than the final secret key that is agreed upon at the end of the QKD cryptographic protocol).

\bigskip

\noindent \textbf{Definition} \textit{12} (\textit{authenticity and confidentiality of bidirectional Quantum channels between Alice and Bob}, \textit{1.2} [16]). A Quantum channel shared by Alice and Bob is said to be bidirectionally authentic, and confidential if,

{\small \begin{align*}
\textbf{P} \big[  \textit{Alice transmits a message to Bob that is authentic over the Quantum channel, or} \\ \textit{vice versa}     \big]  = 1 , \\ \\  \textbf{P}  \big[  \textit{Only Alice and Bob can view messages transmitted over the Quantum channel}     \big] \\  = 1   , 
\end{align*} } 

\noindent respectively.

\bigskip

\noindent \textbf{Definition} \textit{13} (\textit{the 0-entropy from the min-entropy}, \textit{1.1} [16]). Fix two random variables $X$ and $Y$ with ranges $\mathcal{X}$ and $\mathcal{Y}$, respectively. Denote the $0$-entropy as $H_0 \big[ Y \big] : = \mathrm{log} \big[ \big|  y \in \mathcal{Y} : \textbf{P}_Y \big[ y \big] > 0     \big|  \big]$. Besides the $0$-entropy denote the min-entropy as $H_{\infty} \big[ Y \big] : = - \mathrm{log} \big[     \big| \underset{y \in \mathcal{Y}}{\mathrm{sup}} \big\{ \textbf{P}_Y \big[ y \big] \big\}  \big| \big]$.

\bigskip

\noindent \textbf{Definition} \textit{14} (\textit{asymmetric, and symmetric, cryptographic results between the Quantum channel shared by Alice and Bob from the 0, and min, entropies}, \textit{1.2} [16]). One has that asymmetric, and symmetric, cryptographic results, satisfy the following conditions,

\begin{itemize}
    \item[$\bullet$] \textit{(1)}. A cryptographic protocol is said to exhibit \textit{asymmetric} security if, over a Quantum channel shared by Alice and Bob, given an instance $u$ of Eve's random variable $U$, $H_{\infty} \big[ Y \big| \big\{ U = u \big\}  \big] - \underset{x \in \mathcal{X}}{\mathrm{sup}} \big\{  H_0 \big[ Y \big| \big\{ X = x \big\} \big]  \big\} = \Omega \big[ \mathrm{log} \big| \mathcal{Y} \big| \big]$,

    \item[$\bullet$] \textit{(2)}. A cryptographic protocol is said to exhibit \textit{symmetric} security if, over a Quantum channel shared by Alice and Bob, given an instance $u$ of Eve's random variable $U$, $\mathrm{sup} \big\{ H_{\infty} \big[ X \big| \big\{ U = u \big\} \big]  ,  H_{\infty} \big[ Y \big| \big\{ U = u \big\} \big]   \big\} - \underset{x \in \mathcal{X}}{\mathrm{sup}} \big\{  H_0 \big[ Y \big| X  \big]    \big\}  - \underset{y \in \mathcal{Y}}{\mathrm{sup}} \big\{   H_0 \big[ X \big| Y  \big]                \big\} = \Omega \big[ \mathrm{sup} \big\{ \mathrm{log} \big| \mathcal{X} \big|  , \mathrm{log} \big| \mathcal{Y} \big| \big\}    \big] $,
\end{itemize}

\noindent respectively.

\bigskip

\noindent \textbf{Definition} \textit{15} (\textit{two-universal universal functions as alternatives to the two-universal hashing functions for the authentication, information reconciliation and privacy amplification protocols}). Fix $n \neq D > 0$ and some $\epsilon^{\prime}$ sufficiently small. Denote $\mathcal{H} : \big\{ 0 ,1 \big\}^n \longrightarrow \big\{ 0 , 1 \big\}^D$ as the two-universal function with the collision bound,

\begin{align*}
  \textbf{P} \big[ \textit{random variables } x \neq x^{\prime}   :  \mathcal{H} \big( x \big) = \mathcal{H} \big( x^{\prime} \big)  \big] \leq \epsilon^{\prime}   .
\end{align*}

\noindent \textbf{Definition} \textit{16} (\textit{authentication, information reconciliation and privacy amplification protocols for asymmetric unconditional security}, \textit{2.1} [16]). Fix $H_A \neq H_B \in \mathcal{H}$ corresponding to the Alice's two-universal function and Bob's two-universal function, respectively supported over their Hilbert spaces $\mathcal{H}_A$ and $\mathcal{H}_B$. Also, fix $X_A$ and $X_B$ corresponding to two random variables from Alice and Bob, respectively. Introduce,

\begin{align*}
  \pi_{\mathrm{AUTH}} \sim \mathcal{H}_A \times \mathcal{H}_B , \\ \\ \pi_{\mathrm{IR}} \sim X_A \times X_B , \\ \\  \pi_{\mathrm{PA}}   \sim   X_A \times X_B  , 
\end{align*}

\noindent with,

{\small \begin{align*}
  \pi_{\mathrm{AUTH}} \equiv   \big[ \pi_{\mathrm{AUTH}}  \big|_A ,  \pi_{\mathrm{AUTH}}\big|_B \big] =     \big[ \textbf{P} \big[ \textit{Alice authenticates a message that she receives from Bob} \big] \\  , \textbf{P} \big[ \textit{Bob transmits a message to Eve} \big] \big]        , \end{align*}
  \begin{align*} \pi_{\mathrm{IR}}    \equiv      \big[ \pi_{\mathrm{IR}}  \big|_A ,  \pi_{\mathrm{IR}}\big|_B \big] =    \big[ \textbf{P} \big[                     \textit{With $\pi_A$ Alice uses $H_A$ to determine whether the output of her function } \\ \textit{on $X_A$ agrees with the output of $H_B$ on $X_B$}      \big]  , \textbf{P} \big[       \textit{With $\pi_B$ Bob uses $H_B$ to determine whether the } \\ \textit{output of his function on $X_B$ agrees with the output of $H_A$ on $X_A$}                \big]  \big]             , \\ \\     \pi_{\mathrm{PA}}   \equiv       \big[ \pi_{\mathrm{PA}}  \big|_A ,  \pi_{\mathrm{PA}}\big|_B \big] =       \big[ \textbf{P} \big[  \textit{Alice decreases the probability that Eve can reveal any of the bits across} \\ \textit{  all blocks of $X_A$}    \big] , \textbf{P} \big[    \textit{Bob decreases the probability that Eve can reveal any of the bits across} \\ \textit{  all blocks of $X_B$}    \big]  \big]                 , 
\end{align*} }

\noindent corresponding to the policies of Alice and Bob for the authentication, information reconciliation and privacy amplification protocols, respectively.

\bigskip

\noindent We conclude the section with an alternative definition of the min-entropy in terms of the guessing probability for Eve.

\bigskip

\noindent \textbf{Definition} \textit{17} (\textit{alternative formulation of the min-entropy given Eve's side information $E$}). Given Eve's side information $E$ about $X_A$ or $X_B$, the min-entropy can also be defined in terms of the natural logarithm of $\big[ \textbf{P}_{\mathrm{guess}} \big[ X \big| E \big]  \big]^{-1}$.

\subsection{Statement of main results}

We introduce the following objects for stating the main results.

\bigskip

\noindent \textbf{Definition} \textit{18} (\textit{joint probability distribution over candidate Quantum keys from Alice and Bob}). Denote $\textbf{P}_{X_A X_B} \big[ \cdot \big]$ as the joint probability distribution over the product of Quantum secret keys constructed by Alice and Bob, respectively.

\bigskip

\noindent \textbf{Definition} \textit{19} (\textit{side information that Eve gathers from communication between Alice and Bob over the noisy Quantum channel}). Denote $E$ as the information that Eve gathers from Alice's and Bob's communication over the noisy Quantum channel.

\bigskip

\noindent To ensure that Alice and Bob can extract a shared Quantum secret key through weak correlated randomness over an insecure public channel, crucially we make use of the following upper bound from the Holevo information.

\bigskip

\noindent \textbf{Definition} \textit{20} (\textit{upper bound from the Holevo information from the min-entropy conditioned on Eve's side information}). An upper bound from the Holevo information is obtained from an inequality of the form,

\begin{align*}
    \underset{\textit{measurements over blocks of $X_A$ and $X_B$}}{\mathrm{sup}} \big\{  \textbf{P} \big[ \textit{Eve guesses $X_A$ or $X_B$} \big]  \big\}  \leq  2^{-H_{\infty} \big[ X \big| E \big]}  . 
\end{align*}

\bigskip

\noindent \textbf{Definition} \textit{21} (\textit{completely positive trace preserving maps}). A map is said to be CPTP if it satisfies the following properties:

\begin{itemize}
    \item[$\bullet$] \textit{CPTP positivity}. The image of the CPTP map into $\mathcal{B} \big[ \mathcal{H}_A \big]$ is strictly positive iff the input probability itself is strictly positive.

\bigskip

\item[$\bullet$] \textit{CPTP complete positivity}. A CPTP map is completely positive iff the tensor product of the CPTP map with the identity operator is positive.

    \noindent

    \bigskip

     \item[$\bullet$] \textit{CPTP linearity}. One has that,

     \begin{align*}
  \Phi \big[ \alpha \rho + \beta \sigma  \big] =  \alpha \Phi \big[ \rho \big]  + \beta  \Phi \big[  \sigma  \big]                ,
     \end{align*}

     \noindent for $\alpha, \beta \in \textbf{C}$, and,

       \begin{align*}
      \rho \equiv  \textit{State distributed to Eve with support over $\mathcal{H}_A$}  , \\ \\ \sigma \equiv \textit{State distributed to Eve with support over $\mathcal{H}_B$} . 
    \end{align*}

     \bigskip

      \item[$\bullet$] \textit{CPTP tensor product closure}. Denote $\Phi^{\prime}$ as another CPTP which is not equal to $\Phi$. Then $\Phi^{\prime} \otimes \Phi$ is also CPTP.

\bigskip

      \item[$\bullet$] \textit{CPTP composition closure}. Denote $\Phi^{\prime}$ as another CPTP which is not equal to $\Phi$. Then the composition $\Phi^{\prime} \circ \Phi$ is also CPTP.

      \bigskip

       \item[$\bullet$] \textit{CPTP contractivity with respect to the trace norm}. Eve's error probability is shown to satisfy the above inequality with respect to the binary entropy function as a result of the \textit{CPTP contractivity property}, which states,

       \begin{align*}
        \big| \big| \Phi \big[ \rho \big] - \Phi \big[ \sigma \big] \big| \big|_1 \leq \big| \big| \rho - \sigma \big| \big|_1    .
        \end{align*}

        \end{itemize}

\bigskip

\noindent \textbf{Definition} \textit{22} (\textit{Eve's tags for her attempts at forgery over the Quantum channel through a conditionally defined probability}). Take real $\lambda$ sufficiently small. If Alice or Bob transmit a message $M$ over the Quantum channel, denote $T$ as the tag that is generated by $X_A$ or $X_B$. For Eve to attempt a forgery of $M$, she must not only generate a counterfeit tag $T^{\prime} \neq T$, but also some $M^{\prime} \neq M$ for which her false acceptance probability satisfies,

\begin{align*}
  p_{\mathrm{FA}, \mathrm{E}} = \textbf{P} \big[ \textit{Bob mistakenly accepts $\big( M^{\prime} , T^{\prime} \big)$ from Eve} \big| M \neq M^{\prime} \big] \leq 2^{-\lambda}  .
\end{align*}

\noindent Fix $M_A \neq M_B >0$. We make use of the following observations for upper bounding $ \underset{\text{measurements}}{\mathrm{sup}}  \textbf{P} \big[ \textit{Eve guesses $X_A$ or $X_B$} \big] $. Specifically, from previous work of the author which characterized CPTP maps for composable, unconditional security, [22],

\begin{align*}
    \underset{\text{measurements}}{\mathrm{sup}} \big\{  \textbf{P} \big[ \textit{Eve guesses $X_A$ or $X_B$} \big] \big\} \leq f_{\mathrm{DP}} \big[ \chi_E \big[  X_A \textit{ or } X_B  \big] \big]
    \end{align*}

\noindent where $F_{\mathrm{DP}} \big[ \cdot \big]$ is a function obtained from the data-processing inequality,

 \begin{align*}
      D \big[ \rho \big| \big| \sigma \big] \geq   D \big[ \Phi \big[ \rho \big] \big| \big| \Phi \big[ \sigma \big] \big]     , 
 \end{align*}

\noindent of two states $\rho$ and $sigma$, in addition to $\chi_E \big[ X_A \textit{ or } X_B  \big]$,

 \begin{align*}
       \chi_E \equiv  S \big[ \bar{\sigma_A} \textit{ or } \bar{\sigma_B} \big] - \frac{1}{M_A \textbf{1}_{\{\chi_E \equiv \chi_E [ X_A ] \} }  + M_B \textbf{1}_{\{ \chi_E \equiv \chi_E [ X_B ] \} }} \underset{u \in \mathcal{H}}{\sum} S \big[ \sigma_A \textit{ or } \sigma_B \big]     , 
 \end{align*}

\noindent is Eve's Holevo information about $X_A$, or $X_B$ for the von Neummann entropy $S \big[ \cdot \big]$.

\bigskip

\noindent Altogether, with Eve's Holevo information we relate the Holevo information to the min entropy by arguing that the Rennner-Wolf condition on the min-entropy provided in [16], $H_{\infty} \big[ X \big| E \big] \gtrsim n$ implies that $\chi_E \big[ X_A \textit{  or } X_B \big] \leq \delta$, where $n \neq \delta$ are each taken sufficiently small. In the following result below, after having converted conditions from the min-entropy to the Holevo information we make use of a universal function sampled independently from $\mathcal{H}$ without having to analyze every interactive step of the Authentication, Information Reconciliation and Privacy Amplification protocols provided in [16].

In the main result below we we present a generalized framework allowing for: (1) reformulating conditions from the min entropy with the Holevo information; (2) incorporating previous arguments due to the author pertaining to explicit bounds for Eve's false acceptance probability; (3) establishing modularity in the protocol, as a result treating authentication as a resource for Alice and Bob; (4) integrating the steps of interactive authentication from [16] into a single, unified security parameter which is dependent upon the data-processing function evaluated at Eve's Holevo information of $X_A$ or $X_B$, as well as upon Eve's false acceptance probability.

\bigskip

\noindent \textbf{Theorem} (\textit{Holevo information of weak correlated randomness between Alice and Bob and a unified security threshold}). Let Alice and Bob share weak correlated randomness through the candidate Quantum secret keys $X_A$ and $X_B$, which Eve gathers information about through $E$. Given an authentication tag $T$ that Alice or Bob produce when generating $X_A$, or $X_B$, respectively, there exists a protocol that is composed of the following steps:

\begin{itemize}
    \item[$\bullet$] \textit{Error correction}. Given the candidate secret keys $X_A$ and $X_B$, the probability that Eve can learn a key for forging over the noisy Quantum channel has the upper bound $f_{\mathrm{DP}} \big[ \chi_{E,C} \big[ X_A \textit{ or } X_B \big] \big] + p_{\mathrm{FA},\mathrm{E}}$.

    \bigskip

        \item[$\bullet$] \textit{Privacy amplification}. To decrease Eve's forgery probability as much as possible, Alice and Bob can implement a privacy amplification protocol, for the purposes of obtaining composable security.

\bigskip

            \item[$\bullet$] \textit{Authentication based on hashing from two-universal functions}. Fix a secrecy threshold $\epsilon_S \neq 0$ taken sufficiently small. Alice and Bob can authenticate that the secret key that they wish to agree upon has length $l >0$ satisfying,

                \begin{align*}
                    l   \lesssim n - \chi_E \big[ X_A \textit{ or } X_B] - \mathrm{log} \big[ \epsilon^{-1}_S \big]      .
                \end{align*}
\end{itemize}

\noindent Hence the interactive steps of the interactive protocol described in [16] are formalized through a single security threshold $\epsilon \equiv f_{\mathrm{DP}} \big[     \chi_{E,C} \big[ X_A \textit{ or } X_B \big]   \big] + p_{\mathrm{FA},\mathrm{E}}$. 

\bigskip

\noindent To argue that Eve's false acceptance probability upper bounds her forgery probability, in addition to contributions in the upper bound from the function in the Data-Processing inequality, we make use of the following security analysis:



{\small \[ \left\{\!\begin{array}{ll@{}l} 
  \textit{Upper bounding Eve's guessing probability of $M$ with the function from the Data-Processing inequality}. 
\\  \text{One has that }  p_E \leq f_{\mathrm{DP}} \big[ \chi_E \big[ X_A \textit{ or } X_B \big] \big].
\\  \\  \textit{Eve's false acceptance probability can be made arbitrarily small}. \text{ One has that } p_{\mathrm{FA}, \mathrm{E}} \leq 2^{-\lambda} \text{ for }\lambda \text{ taken} \\ 
 \text{sufficiently large}. \\ \\ \textit{Composability of unconditional security between Alice and Bob}. \text{ Given the analysis of [22], if Eve's guessing} \\ \text{and false acceptance probability are arbitrarily small, then the protocol between Alice and Bob for gen-} \\ \text{erating a Quantum secret key of length $l$ that they agree upon - Quantum secret key extraction - is} \\ \text{composable.} 
\end{array}\right. \]  }

\noindent We denote the above analysis with $(*)$. To demonstrate that \textbf{Theorem} holds, we make use of the following result.

\bigskip

\noindent \textbf{Lemma} (\textit{upper bounding Eve's guessing probability with the Relative min-entropy and her Holevo information of $X_A$ or $X_B$}). Fix $X_A \neq X_B \in \big\{ 0 , 1 \big\}^n$ corresponding to random variables in Quantum secret keys held by Alice and Bob, respectively and $\delta >0$ sufficiently small. Suppose that, $\chi_E \big[ X_A \textit{ or } X_B \big] \leq \delta$, and also that, $p_{\mathrm{guess}} \big[ X_A \textit{ or } X_B \big| E \big] \leq 2^{-H_{\infty} [ X_A \textit{ or } X_B \big| E ] }$. With the Relative-min entropy and Eve's Holevo information of $X_A$ or $X_B$, her guessing and forgery probabilities satisfy,

\begin{align*}
   p_{\mathrm{forge}}   \leq  p_{\mathrm{guess}} \big[ X_A \textit{ or } X_B \big| E \big] \leq 2^{-         H [ X_A \textit{ or } X_B ] - \delta      } . 
\end{align*}

\noindent As a result $H_{\infty} \big[ X_A \textit{ or } X_B \big| E \big] \geq H \big[ X_A \textit{ or } X_B \big] - \delta$.

\bigskip

\noindent We present additional results related to the \textbf{Theorem} and \textbf{Lemma} stated above. Crucially, given the fact that the security thresholds associated within the Renner-Wolf framework in [16] are dependent upon the observation that,

\begin{align*}
    \textbf{P} \big[ \textit{Eve's forgery is successful} \big]  \leq 2^{-H_{\infty} [ X_A \textit{ or } X_B | E ]}       , 
\end{align*}

\noindent one can instead bound her forgery probability with the Holevo information through, 

\begin{align*}
     \textbf{P} \big[ \textit{Eve commits a forgery} \big] \leq 2^{- [ H [ X_A \textit{ or } X_B ] - \chi_E [ X_A \textit{ or }  X_B ] ]}  .
\end{align*}

\noindent Below we state how one can characterize the existence of a security threshold with either positive, or zero, probability.

\bigskip

\noindent \textbf{Theorem} \textit{2} (\textit{the Holevo-gap threshold}). If Alice generates $X_A$ with Bob who wants to generate $X_B$ that they agree upon if $X_A \approx X_B$ at the end of the protocol, from the Holevo gap,

\begin{align*}
  \Delta \equiv H \big[ X_A \textit{ or } X_B \big] - \chi_E \big[ X_A \textit{ or } X_B \big]  , 
\end{align*}

\noindent one has that:

\begin{itemize}
    \item[$\bullet$] Suppose that $\Delta >0$. There exists an information-theoretic protocol with the corresponding security threshold $2^{-\Delta}$ with positive probability.

    \item[$\bullet$] Suppose that $\Delta \leq 0$. There exists an information-theoretic protocol with the corresponding security threshold $2^{-\Delta}$ with vanishing probability, and thus, $p_{\mathrm{forge}} \geq \Omega \big[ 1 \big]$.
\end{itemize}

\noindent Essentially the Holevo gap threshold determines whether an authenticated channel can be realized. It is also of interest to determine whether a converse result which is dependent upon the Holevo gap can be used to lower bound Eve's forgery probability.

\bigskip

\noindent Relatedly, one can upper bound the authentication probability in terms of the authentication protocol from a weakly random source.

\bigskip

\noindent \textbf{Corollary} (\textit{upper bounding the authentication probability from the Holevo information}). Fix some $k$ sufficiently small. For the authentication protocol in [16] with $X_A$ as the initially shared random secret from Alice that she transmits to Bob, under the assumption,

\begin{align*}
  \chi_{E,C} \big[ X_A \textit{ or } X_B \big] \leq H \big[ X_A \textit{ or } X_B \big]  - k  ,
\end{align*}

\noindent then,

\begin{align*}
   p_{\mathrm{auth}}   \leq 2^{-k}    . 
\end{align*}

\noindent We also state how the length of the extracted key that Alice and Bob agree is dependent upon the Holevo information.

\bigskip

\noindent \textbf{Corollary} \textit{2} (\textit{the extracted key length and the Holevo information}). Fix $l > k >0$. If $\chi_E \big[ X_A \textit{ or } X_B \big] \leq H \big[ X_A \textit{ or } X_B \big] - k - l $, then there exists a protocol for which:

\begin{itemize}
    \item[$\bullet$] The probability that Eve makes an error when authenticating $X_A$ or $X_B$ is less than or equal to $2^{-k}$.
    \item[$\bullet$] The probability that Eve extracts a Quantum secret key shared between Alice or Bob, which is dependent upon that probability that $X_A \approx X_B$, with secrecy error, is less than or equal to $2^{-l}$.
\end{itemize}

\noindent We state the final result below pertaining to a leakage bound from the Holevo information.

\bigskip

\noindent \textbf{Lemma} \textit{2} (\textit{stability under public discussion via the Holevo information}). Denote $C$ as the transcript of public communication over the noisy Quantum channel between Alice and Bob of $t >0$ bits. One has that,

\begin{align*}
  \chi_{E,C} \big[ X_A \textit{ or } X_B \big] \leq \chi_E \big[ X_A \textit{ or } X_B \big] + t   . 
\end{align*}

\noindent With all of the main results stated we present the arguments in the next section.

\section{Arguments}

\noindent We provide arguments for each of the main results stated in the previous section.

\subsection{Lemma}

\noindent \textit{Proof of Lemma}. Fix some $\delta \geq 0$. Suppose $\chi_{E,C} \big[ X_A \textit{ or } X_B \big] \leq \delta$. To argue that,

\begin{align*}
  p_{\mathrm{guess}} \big[ X_A \textit{ or } X_B   \big| E \big]       \leq 2^{-H [ X_A \textit{ or } X_B ] - \delta }  , 
\end{align*}

\noindent and hence that,

\begin{align*}
 H_{\infty}     \big[ X_A \textit{ or } X_B \big| E \big]  \geq H \big[ X_A \textit{ or } X_B \big] - \delta       , 
\end{align*}

\noindent by Holevo's Theorem observe,

\begin{align*}
     I \big[ \big\{ X_A \textit{ or } X_B \big\} , \big\{  \textit{Eve's measurement associated with $X_B$} \big\}  \big] \leq \chi_E \big[ X_A \textit{ or } X_B \big]      . 
\end{align*}

\noindent Furthermore

\begin{align*}
    p_{\mathrm{guess}} \big[ X_A \textit{ or } X_B  \big| E \big]    \leq     2^{-H [ X_A \textit{ or } H_B | \{  \textit{Eve's measurement associated with $X_B$} \} ]}                         . 
\end{align*}

\noindent for an estimator $\hat{X_B \big[ \textit{Eve's measurement associated with $X_B$}   \big] }$ of $X_B \big[ \textit{Eve's measurement associated with $X_B$}  \big]$. The above inequality holds because $H \big[ X_A \textit{ or } X_B \big| \textit{Eve's measurement associated with $X_B$} \big] = H \big[ X_A \textit{ or } X_B \big] - I \big[ \big\{ X_A \textit{ or } X_B \big\} , \big\{  \textit{Eve's measurement associated with $X_B$} \big\} \big]$. Altogether,

{\small \begin{align*}
     p_{\mathrm{guess}} \big[ X_A \textit{ or } X_B \big| \big\{  \textit{Eve's measurement associated with $X_B$} \big\} \big] \leq \frac{2^{-H [ X_A \textit{ or } X_B ]}}{2^{- I [ \{ X_A \textit{ or } X_B \} ,\{  \textit{Eve's measurement associated with $X_B$}\}  ]}}             . 
\end{align*} }

\noindent Taking the supremum over all measurements of,

{\small \begin{align*}
    \underset{\textit{measurements}}{\mathrm{sup}} \big\{ p_{\mathrm{guess}} \big[ \big\{  X_A \textit{ or } X_B \big\} \big| \big\{  \textit{Eve's measurement associated with $X_B$} \big\} \big]   \big\}    ,
\end{align*} } 

\noindent implies,

{\small \begin{align*}
    \frac{2^{-H [ X_A \textit{ or } X_B ]}}{2^{- I [ \{ X_A \textit{ or } X_B \} ,\{  \textit{Eve's measurement associated with $X_B$}\}  ]}}   \leq    \frac{2^{-H [ X_A \textit{ or } X_B ]}}{2^{- \chi_E  [  X_A \textit{ or } X_B   ]}}         , 
\end{align*} } 

\noindent which by the alternative definition of the min-entropy provided in terms of the natural logarithm of Eve's guessing probability provided in \textbf{Definition} \textit{17} yields the desired result, from which we conclude the argument. \boxed{}

\subsection{Theorem}

\noindent \textit{Proof of Theorem}. It suffices to argue that there exists a two-universal function $H \sim \mathcal{H}$ for which,

\begin{align*}
      \textbf{P} \big[ \textit{random variables } x \neq x^{\prime}   :  H  \big( x \big) = H \big( x^{\prime} \big)  \big]        \lesssim p_{\mathrm{FA}, \mathrm{E}}      .
\end{align*}

\noindent Denoting two-universal functions which Alice and Bob respectively use for hashing as $H_A$ and $H_B$, we demonstrate that the composability property of the protocol holds; this property claims that Alice and Bob implement for error correction. In order for Alice and Bob to generate a Quantum secret key that they agree upon which Eve cannot intercept with high probability at the end of the protocol, Alice shares her syndrome,

\begin{align*}
    S_{X_A}  \equiv  \underset{\textit{blocks of } X_A}{\bigcup} \textbf{1}_{\{ \textit{Alice finds a X, or Z, logical error along each block of $X_A$} \}}  ,
\end{align*}

\noindent associated with her candidate Quantum secret key $X_A$ to Bob so that he can reconstruct his candidate Quantum secret key $X_B$ so that $X_A \approx X_A$ whp, namely that, 

\begin{align*}
  \textbf{P} \big[ \textit{Alice and Bob agree that $X_A \approx X_B$ at the end of their cryptographic protocol}  \big] \\ \geq 1 - C_{\epsilon} , 
\end{align*}

\noindent for some $C_{\epsilon}$ taken sufficiently small. After having received the syndrome, Alice and Bob can directly apply the authentication, information reconciliation and privacy amplification protocols developed in [16] which are defined probabilistically in \textbf{Definition} \textit{16}; furthermore, equipped with the weaker class of information-theoretic primitives defined in [16], through the computation of the upper bound,

{\small \begin{align*}
  \textbf{P} \big[  X_A , X_B :  \textit{Bob constructs $X_B$ from $X_A$ that he receives from Alice}   \big]  \\ \\ \leq     \textbf{P} \big[  X_A , X_B :  \textit{Bob and Alice agree that $X_A \approx X_B$ after performing inf-}  \\ \textit{ormation reconciliation}   \big] \\ \\ \leq  \textbf{P} \big[  X_A , X_B :  \textit{Bob and Alice agree that $X_A \approx X_B$ after performing inf-}  \\ \textit{ormation reconciliation and privacy amplification} \big]  \\ \\ \leq   \textbf{P} \big[  X_A , X_B :  \textit{Bob and Alice agree that $X_A \approx X_B$ after performing inf-}  \\ \textit{ormation reconciliation, privacy amplification and authentication} \big]     \\ \\  \leq   \textbf{P} \big[  \textit{protocols $\Pi$}:  \textit{Bob and Alice agree that $X_A \approx X_B$ produced by $\Pi$}  \\ \textit{after performing information reconciliation, privacy amp-} \\ \textit{lification and authentication} \big] \\   \\ \approx        \textbf{P} \big[  \textit{protocols $\Pi$}, \textit{two-universal $H_A \neq H_B \sim \mathcal{H}$}: \big\{  \textit{Bob and Alice agree that } \\  \textit{$X_A$ $\approx X_B$}  \textit{ produced by $\Pi$ after performing information reconciliation, pri-} \\ \textit{vacy amplification and authentication} \big\} , \big\{ \big\{  H_A \big[ X \big] \equiv H_B \big[ X^{\prime} \big] \big\}  \Longleftrightarrow \\  \big\{ X \equiv X^{\prime} \big\}       \big\} \big]  \\ \\     \lesssim   \textbf{P} \big[  \textit{protocols $\Pi$} : \big\{  \textit{Bob and Alice agree that $X_A$ $\approx X_B$}  \textit{ produced by $\Pi$ } \\   \textit{after performing information reconciliation, privacy amplification} \\ \textit{and authentication} \big\} ,  \big\{  p_{\mathrm{FA},\mathrm{E}} \textit{ occurs whp} \big\} \big]  \end{align*}}

  {\small \begin{align*} \propto \underset{\mathrm{bits} \longrightarrow + \infty}{\mathrm{lim}} \bigg\{   \textbf{P} \big[  \textit{protocols $\Pi$ dependent upon finitely many bits} : \big\{  \textit{Bob and Alice agree that $X_A$ $\approx X_B$} \\  \textit{ produced} \\ \textit{  by $\Pi$ depending upon finitely many bits after performing information  reconciliation, privacy ampli-}  \\ \textit{  fication and authentication} \big\}  ,  \big\{  p_{\mathrm{FA},\mathrm{E}} \textit{ occurs whp depending upon finitely many bits} \big\} \big]         \bigg\}  \\ \\    \approx \underset{\mathrm{bits} \longrightarrow + \infty}{\mathrm{lim}} \bigg\{     \textbf{P} \big[  \textit{protocols $\Pi$ dependent upon finitely many bits} : \big\{  \textit{Bob and Alice agree that $X_A$ $\approx X_B$}  \\ \textit{ produced} \\ \textit{  by $\Pi$ depending upon finitely many bits after performing information  reconciliation, privacy ampli-}  \\ \textit{  fication and authentication} \big\}     \bigg\} \\ \times  \underset{\mathrm{bits} \longrightarrow + \infty}{\mathrm{lim}} \bigg\{  \textbf{P} \big[   \textit{protocols $\Pi$ dependent upon finitely many bits} :   \big\{  p_{\mathrm{FA},\mathrm{E}} \textit{ occurs whp depending upon} \\ \textit{finitely many bits} \big\} \big]         \bigg\}    \\ \\ \approx     \underset{\mathrm{bits} \longrightarrow + \infty}{\mathrm{lim}} \bigg\{     \textbf{P} \big[  \textit{protocols $\Pi$ dependent upon finitely many bits} : \big\{  \textit{Bob and Alice agree that $X_A$ $\approx X_B$} \\  \textit{ produced} \\ \textit{  by $\Pi$ depending upon finitely many bits after performing information  reconciliation, privacy ampli-}  \\ \textit{  fication and authentication} \big\}     \bigg\}    \\ \times  \underset{\mathrm{bits} \longrightarrow + \infty}{\mathrm{lim}} \bigg\{   \textbf{P} \big[ \textit{random variables } x \neq x^{\prime}  \textit{ depending upon finitely many bits} :  H  \big( x \big) = H \big( x^{\prime} \big)  \big]  \bigg\}    \\ \\   \lesssim  \epsilon  p_{\mathrm{FA},\mathrm{E}}                         , 
\end{align*} }

\noindent one obtains the desired unified security threshold $\epsilon$. With a two-universal function introduced in \textbf{Definition} \textit{15}, Alice can compute an authentication tag associated with $X_A$ and its accompanying message $M$, with,

\begin{align*}
    T_{X_A} \equiv  H_A \big( X_A , M \big)  \sim   \mathcal{H} \big( X_A , M \big)     .
\end{align*}

\noindent Eve's forgery probability is strictly lower bounded by her false acceptance probability, satisfying (\textbf{Theorem} \textit{1}, [22]):

\begin{itemize}

 \item [$\bullet$] \textit{Minimization of probabilities via the Fano inequality}. One has that,

\begin{align*}
  p_{\mathrm{FA},\mathrm{E}} \equiv \underset{\chi_E > 0}{\underset{M > 0}{\bigcup}}  \mathrm{inf} \big\{ p :      \mathrm{log} M - \chi_E \leq  H_b \big(  p \big)   + p \mathrm{log}  \big[ M - 1 \big]          \big\}   .
\end{align*}

\item[$\bullet$] \textit{Up to constant lowers bounds for Eve's error probability}. One has that,

\begin{align*}
  P_{\mathrm{FA},\mathrm{E}} \gtrsim  \mathrm{log} \bigg[ \frac{1}{N} \bigg] \bigg[ \mathrm{log} M - \chi_E - 1 \bigg]  ,
\end{align*}

\noindent for $N \neq M$, each of which can be taken to be sufficiently large.

\bigskip

\item[$\bullet$] \textit{Sharpening the up to constants lower bound for Eve's error probability to an inequality via Helstrom}. One has that,

\begin{align*}
    P_{\mathrm{FA}, \mathrm{E}} \geq 1 - \frac{1 + \epsilon^{\prime\prime} \big( M - 1 \big)}{M}      , 
\end{align*}

\noindent for $\epsilon^{\prime\prime}$ taken to be sufficiently small.

\end{itemize}

\bigskip

\noindent In order for Bob to be able to generate his own tag of authentication for $X_B$, Eve must introduce a suitable decoding scheme over the noisy Quantum channel. After having transmitted the information through the decoding to Bob, straightforwardly his authentication tag is of the form,

\begin{align*}
   T_{X_B} \equiv     H_B \big( X_B , M^{\prime} \big)        \sim \mathcal{H} \big( X_B , M^{\prime} \big) , 
\end{align*}

\noindent where $M^{\prime}$ is the encoding of $M$ transmitted by Alice. Next, by making use of the security analysis provided in $(*)$, Alice and Bob can guarantee that Eve's guessing, and false acceptance, probabilities can be made arbitrarily small.

\bigskip

\noindent Fix some $k>0$. Besides composability of the protocols which Alice and Bob use which will be shown to hold, the assumption provided in [16] on the min-entropy, $H_{\infty} \big[ X_A \textit{ or } X_B \big| E \big] \geq k$. is replaced with the assumption on Eve's Holevo information given her side information, namely $\chi_E \big[ X_A \textit{ or } X_B \big] \leq H \big[ X_A \textit{ or } X_B \big] - k$. With this assumption, apply the data-processing inequality on Eve's Holevo information to conclude,

{\small \begin{align*}
 \big\{  \chi_{E,C} \big[ X_A \textit{ or } X_B \big] \leq \chi_E \big[ X_A \textit{ or } X_B ] + l_{\textit{leakage}}       \big\} \Longrightarrow \big\{       H_{\infty} \big[  X_A \textit{ or } X_B \big] \geq H \big[  X_A \textit{ or } X_B  \big] - \chi_{E,C} \big[  X_A \\ \textit{ or } X_B   \big]   \big\}    ,
\end{align*} }

\noindent for $l_{\textit{leakage}} > 0$.

\bigskip

\noindent To demonstrate that the unified security parameter satisfies a composability property as characterized in previous works of the author, [17,18,19,20,21,22], it suffices to demonstrate that there exists a sequence of authentication tags related to $T_{X_A}$ and $T_{X_B}$, respectively. To this end, observe,

{\small \begin{align*}
   \textbf{P} \big[   T_{X_A} , T_{X_B}:  \textit{Alice and Bob agree that $X_A \approx X_B$ at the end of the cryptographic protocol}          \big]  \\ \\ \leq        \textbf{P} \big[   T_{X_A} , T_{X_B}: \big\{  \textit{Alice and Bob agree that $X_A \approx X_B$ at the end of the cryptographic protocol} \big\} , \\   \big\{   \textit{Alice and Bob agree that $X_A \approx X_B$ after having performed information reconciliation}   \big\}               \big] \\ \\  \leq        \textbf{P} \big[   T_{X_A} , T_{X_B}: \big\{  \textit{Alice and Bob agree that $X_A \approx X_B$ at the end of the cryptographic protocol} \big\} , \\   \big\{   \textit{Alice and Bob agree that $X_A \approx X_B$ after having performed information reconciliation}   \big\} , \\ \big\{   \textit{Alice and Bob agree that $X_A \approx X_B$ after having performed privacy amplification}       \big\}              \big]   \\ \\   \leq        \textbf{P} \big[   T_{X_A} , T_{X_B}: \big\{  \textit{Alice and Bob agree that $X_A \approx X_B$ at the end of the cryptographic protocol} \big\} , \\   \big\{   \textit{Alice and Bob agree that $X_A \approx X_B$ after having performed information reconciliation}   \big\} , \\ \big\{   \textit{Alice and Bob agree that $X_A \approx X_B$ after having performed privacy amplification}       \big\}  , \\ \big\{  \textit{Alice and Bob agree that $X_A \approx X_B$ after having performed authentication}          \big\}             \big]    \\ \\     \overset{(*)}{\leq}       \textbf{P} \big[   T^{\prime}_{X_A} , T^{\prime}_{X_B}: \big\{  \textit{Alice and Bob agree that $X_A \approx X_B$ at the end of the cryptographic protocol} \big\} , \\   \big\{   \textit{Alice and Bob agree that $X_A \approx X_B$ after having performed information reconciliation}   \big\} , \\ \big\{   \textit{Alice and Bob agree that $X_A \approx X_B$ after having performed privacy amplification}       \big\}  , \\ \big\{  \textit{Alice and Bob agree that $X_A \approx X_B$ after having performed authentication}          \big\}             \big]   \\ \vdots \\       \leq       \textbf{P} \big[   T^{\prime\cdots\prime}_{X_A} , T^{\prime\cdots\prime}_{X_B}: \big\{  \textit{Alice and Bob agree that $X_A \approx X_B$ at the end of the cryptographic protocol} \big\} , \\   \big\{   \textit{Alice and Bob agree that $X_A \approx X_B$ after having performed information reconciliation}   \big\} , \\ \big\{   \textit{Alice and Bob agree that $X_A \approx X_B$ after having performed privacy amplification}       \big\}  , \\ \big\{  \textit{Alice and Bob agree that $X_A \approx X_B$ after having performed authentication}          \big\}             \big]    \\ \\ \approx \underset{\mathrm{bits} \longrightarrow + \infty}{\mathrm{lim}}       \textbf{P} \big[   T_{X_A} , T_{X_B}: \big\{  \textit{Alice and Bob agree that $X_A \approx X_B$ at the end of the cryptographic pro-} \\ \textit{tocol} \big\} ,   \big\{   \textit{Alice and Bob agree that $X_A \approx X_B$ after having performed information reconcili-} \\ \textit{ation}   \big\} ,  \big\{   \textit{Alice and Bob agree that $X_A \approx X_B$ after having performed pri-} \\ \textit{vacy amplification}       \big\}  ,  \big\{  \textit{Alice and Bob agree that $X_A \approx X_B$} \\ \textit{ after having performed authentication}          \big\}             \big]    \\ \\                       \lesssim     \big| T_{X_A} \big| \times \cdots \times    \big| T^{\prime\cdots\prime}_{X_A} \big|    \big| T_{X_B} \big| \times \cdots \times    \big| T^{\prime\cdots\prime}_{X_B} \big|         \epsilon  p_{\mathrm{FA},\mathrm{E}}  \\ \\  \lesssim   C \big( X_A, X_B, \epsilon \big) p_{\mathrm{FA},\mathrm{E}}                 .
\end{align*} }

\noindent In the sequence of upper bounds obtained first in $(*)$ between the authentication tags $T_{X_A}$ and $T^{\prime}_{X_A}$, and also between $T_{X_B}$ and $T^{\prime}_{X_B}$, one makes use of the fact that there exists hashing functions supported over messages with infinitely many bits so that the series of authentication tags $\underset{\mathrm{bits} \longrightarrow + \infty}{\mathrm{lim}} \big\{ T^{\prime}_{X_A} \big\}_{1 \leq \prime \leq \mathrm{bits}} $ and $\underset{\mathrm{bits} \longrightarrow + \infty}{\mathrm{lim}} \big\{ T^{\prime}_{X_B} \big\}_{1 \leq \prime \leq \mathrm{bits}} $ have the respective collision bounds,

{\small \begin{align*}
   \textbf{P} \big[ \textit{random variables } X_A  \neq X^{\prime}_A ,  H_A \big( X_A \big)   \sim \mathcal{H} \big( X_A  \big) , H_A \big( X^{\prime}_A \big)   \sim \mathcal{H} \big( X^{\prime}_A  \big)     : H_A \big( X_A \big)  = H_A \big( X^{\prime}_A \big)    \big] \\ \leq \epsilon_A      , \\ \\   \textbf{P} \big[ \textit{random variables } X_B \neq X^{\prime}_B , H_B \big( X_B \big)  \sim \mathcal{H} \big( X_B  \big) ,   H_B \big( X^{\prime}_B \big)  \sim  \mathcal{H} \big( X^{\prime}_B  \big)   :  H_B \big( X_B \big)  = H_B \big( X^{\prime}_B \big) \big] \\ \leq \epsilon_B       . \\ 
\end{align*} }

\noindent for $\epsilon_A \neq \epsilon_B$, each of which are taken to be sufficiently small.

\bigskip

\noindent We return to arguments for lower bounding Eve's false acceptance probability with the probability of sampling two random variables related to the criteria $\big\{ \textit{random variables } x \neq x^{\prime}   :  H  \big( x \big) = H \big( x^{\prime} \big)  \big\}$. The following computation, in addition to the previous one which obtained the up to constants estimate $C \big( X_A , X_B, \epsilon \big) p_{\mathrm{FA},\mathrm{E}}$,  implies that the length $l$ of the extracted Quantum secret key shared between Alice and Bob satisfies the desired up to constants estimate hence demonstrating that the first main result holds. To this end, if one denotes,

\begin{align*}
    \textbf{P} \big[ \textit{Eve commits an instance of false acceptance} \big]   \equiv     \textbf{E} \big[  \textbf{P} \big[  \textit{Eve commits a forgery}        \big| \big\{ \big\{ x_A   \sim X_A \big\}  , \\ \big\{  x_B  \sim X_B  \big\}  \big\}  , \big\{ e \sim E  \big\} \big]   \big]           ,
\end{align*}

\noindent for the expectation taken $\textbf{E} \big[ \cdot \big]$ with respect to $\textbf{P} \big[ \cdot \big]$, the observation

\begin{align*}
     \textbf{P} \big[ \textit{Eve commits an instance of false acceptance} \big]    \equiv p_{\mathrm{FA},\mathrm{E}} \leq 2^{-l}                   . 
\end{align*}

\noindent implies, $p_{\mathrm{FA},\mathrm{E}} \leq p_{\mathrm{guess}} \big[ X_A \textit{ or } X_B \big| E \big] + \big[ 1 - p_{\mathrm{guess}} \big[ X_A \textit{ or } X_B \big| E \big] \big]  2^{-l}$, and furthermore, that,

{\small \begin{align*}
      \frac{\textbf{P} \big[ \textit{random variables } x \neq x^{\prime}   :  H  \big( x \big) = H \big( x^{\prime} \big)  \big]}{p_{\mathrm{FA}, \mathrm{E}} }             \equiv                       \textbf{P} \big[ \textit{random variables } x \neq x^{\prime}   :  H  \big( x \big) = H \big( x^{\prime} \big)  \big]    \\ \times   \big\{  \textbf{E} \big[  \textbf{P} \big[  \textit{Eve commits a forgery}        \big| \big\{ x_A   \sim X_A  , x_B  \sim X_B  \big\}  , \big\{ e \sim E  \big\} \big]   \big] \big\}^{-1}  \\           \\     \equiv                        \frac{\textbf{P} \big[ \forall \textit{ messages } M \neq M^{\prime}, \exists \textit{ random variables } \big( x , M \big)  \neq  \big( x , M^{\prime} \big)   :  H  \big( x \big) = H \big( x^{\prime} \big)  \big]}{      \textbf{E} \big[  \textbf{P} \big[  \textit{Eve commits a forgery}        \big| \big\{ x_A   \sim X_A , x_B  \sim X_B  \big\}  , \big\{ e \sim E  \big\} \big]   \big] }                           \\ \\   \overset{(\textit{two-universality of $H \sim \mathcal{H}_A, \mathcal{H}_B$})}{\leq}            \frac{2^{-l}}{      \textbf{E} \big[  \textbf{P} \big[  \textit{Eve commits a forgery}        \big| \big\{ x_A   \sim X_A , x_B  \sim X_B  \big\}  , \big\{ e \sim E  \big\} \big]   \big] }    \\ \\     \leq            \frac{2^{-l}}{    \textbf{E} \big[  \textbf{P} \big[  \textit{Eve commits a forgery}        \big| \big\{ x_A   \sim X_A , x_B  \sim X_B  \big\}   \big]   \big]  }            \\ \\   \lesssim  \frac{C_1 \big( l \big) }{C_2 \big( X_A , X_B \big)  }     \\ \\   \lesssim  C_{1,2} \big( l , X_A , X_B \big) \equiv C_{1,2}       \\ \\ \leq 1   \\ \\        \lesssim    1                 ,
\end{align*}    } 

\noindent from the fact that $p_{\mathrm{FA},\mathrm{E}} \approx p_{\mathrm{guess}} \big[ X_A \textit{ or } X_B \big| E \big]$ for $l$ taken sufficiently large. Altogether,

{\small \begin{align*}
    \textbf{P} \big[ \textit{random variables } x \neq x^{\prime}   :  H  \big( x \big)  = H \big( x^{\prime} \big)  \big]        \lesssim p_{\mathrm{FA}, \mathrm{E}}          .  \\ 
\end{align*} }

 \noindent The above product threshold, the first component of which is dependent upon the image of the data-processing function with the Holevo information, and the second of which is dependent upon the choice of $\lambda$ taken sufficiently large, imply that the security of the protocol that Alice and Bob use for agreeing that $X_A \approx X_B$ whp is composable, hence guaranteeing that the security parameter, by virtue of being composable, holds. As described previously the form of the unified security threshold, in comparison to multiple security thresholds in interactive authentication protocols, is dependent upon the function obtained from the data-processing inequality and Eve's false acceptance probability, from which we conclude the argument. \boxed{} 


\subsection{Theorem $2$}

\noindent \textit{Proof of Theorem 2}. Given the Holevo gap $\Delta$, it suffices to argue that from the first item which claims that a protocol exists, with positive probability, such that a security threshold of $2^{-\Delta}$ holds. Hence observe,

\begin{align*}
       H_{\infty} \big[ X_A \textit{ or } X_B \big| E \big] \geq H \big[ X_A \textit{ or } X_B \big] - \chi_{E} \big[ X_A \textit{ or } X_B \big]       , 
\end{align*}

\noindent implies the desired result, as a result also guaranteeing that,

\begin{align*}
  \textbf{P} \big[  \textit{QKD protocol} : \textit{protocol has security $2^{-\Delta}$}     \big] > 0   . 
\end{align*}

\noindent For the second result which states that the converse of the first result holds, observe that $\chi_E \big[ X_A \textit{ or } X_B \big] \geq H \big[ X_A \textit{ or } X_B \big]$ implies,

\begin{align*}
   \textbf{P} \big[ \textit{candidate Quantum keys $X_A$, $X_B$} : \textit{Eve can reconstruct $X_A$ or $X_B$} \big] \approx 0  . 
\end{align*}

\noindent Altogether $p_{\mathrm{forge}} \geq \Omega \big[ 1 \big]$, from which we conclude the argument. \boxed{}

\subsection{Corollary}

\noindent \textit{Proof of Corollary}. To demonstrate that the authentication probability can be made arbitrarily small for $k$ taken sufficiently large, observe that the assumption $H_{\infty} \big[ X_A \textit{or } X_B \big| E  \big] \geq k$ on the min-entropy implies,

\begin{align*}
  p_{\mathrm{auth}} \leq 2^{-H_{\infty} [ X_A \textit{ or } X_B \big| E]} \leq 2^{-k}  ,
\end{align*}

\noindent from which we conclude the argument. \boxed{}

\subsection{Corollary $2$}

\noindent \textit{Proof of Corollary 2}. Observe from previous results that there exists $k > l > 0$ such that $H_{\infty} \big[ X_A \textit{ or } X_B \big] \geq k + l$. Hence on can demonstrate that the probability of encountering errors for both error and secrecy can be bound simultaneously with $2^{-k} + 2^{-l}$. Moreover if one denotes $k$ as the number of bits of entropy dedicated to authentication and $l$ as the number of bits dedicated to privacy amplification, the $\leq$ bound for the $2^{-k}$ factor is obtained from a direct application of the authentication protocol in [16]. The remaining factor of $2^{-l}$ for extracting the desired key of length $l>0$ after Alice and Bob perform privacy amplification is obtained from the up to constants upper bound,

 \begin{align*}
                    l   \lesssim n - \chi_E \big[ X_A \textit{ or } X_B] - \mathrm{log} \big[ \epsilon^{-1}_S \big]      ,
                \end{align*}

\noindent provided in the third item stated in \textbf{Theorem} implies that the desired contribution of $2^{-l}$ also holds, from which we conclude the argument. \boxed{}

\subsection{Lemma $2$}

\noindent \textit{Proof of Lemma 2}. From direct computation,

\begin{align*}
   I_{E,C} \big[ X_A \textit{ or } X_B \big] =  I_{E,C} \big[    X_A \textit{ or } X_B \big] + I_E \big[ X_A \textit{ or } X_B \big|  C \big] \leq      \chi_E \big[ X_A \textit{ or } X_B \big]    +     H \big[ C \big] \leq     \chi_E \big[ X_A \\ \textit{ or }  X_B \big]   + t           , 
\end{align*}

\noindent by the chain rule of the Mutual Information entropy and the data-processing inequality we conclude the argument. \boxed{}

\section{Conclusion}

\noindent In this work we characterized the security threshold of a cryptographic protocol that shares similarities with the interactive authentication protocol introduced in [16]. In place of assumptions on the min-entropy we made use of several properties of the Holevo information that have previously been analyzed by the author for bounding Eve's false acceptance probability over a noisy Quantum channel. However, in previous work of the author connections between the security threshold of a unified interactive authentication protocol analyzed in this work had not yet been related to the false acceptance probability. Along these lines, we motivated the security threshold obtained in this work as being related to: (1) optimality, and approximate optimality, of Quantum strategies in game-theoretic settings; (2) hashing protocols associated with authentication in QKD cryptographic protocols; (3) applications of the Fano and data processing inequalities for obtaining estimates on Eve's false acceptance probability, which inform estimates on Eve's forgery probability. It is of interest to characterize other sources for prospective Quantum advantage for closely related protocols.

\section{References}


\bigskip


\noindent [1] Gidi, J.A. and Candia, B. and Munoz-Moller, A.D. and Rojas, A. and Pereira, L. and Munoz, M. and Zambrano, L. and Delgado, A. Stochastic optimization algorithms for quantum applications. \textit{Phys.Rev.A} 108: 032409 (2023). https://doi.org/10.1103/PhysRevA.108.032409. 



\bigskip

\noindent [2] Givi, P. and Daley, A.J. and Mavriplis, D. and Malik, M. Quantum Speedup for Aeroscience and Engineering. \textit{AIAA} 58:8 (2020). 

https://ntrs.nasa.gov/api/citations/20200003505/downloads/20200003505.pdf.


\bigskip

\noindent [3] Helton, J.W., Mousavi, H., Nezhadi, S.S. et al. Synchronous Values of Games. \textit{Ann. Henri Poincaré} \textbf{25}, 4357–4397 (2024). https://doi.org/10.1007/s00023-024-01426-1

\bigskip

\noindent [4] Hadiashar, S.B. and Nayak, A. and Sinha, P. Optimal lower bounds for Quantum Learning via Information Theory. \textit{IEEE Transactions on Information Theory} 70(3): 1876--1896 (2024). https://doi.org/10.1109/TIT.2023.3324527. 



\bigskip

\noindent [5] Hur, T. and Kim, L. and Park, D.K. Quantum convolutional neural network for classical data classification. \textit{Quantum Machine Intelligence} 4: 3 (2022). https://doi.org/10.1007/s42484-021-00061-x. 



\bigskip

\noindent [6] Holmes, Z. and Coble, N.J. and Sornborger, A.T. and Subasi, Y. On nonlinear transformations in quantum computation. \textit{Phys. Rev. Research} 5: 013105 (2023). https://doi.org/10.1103/PhysRevResearch.5.013105. 



\bigskip

\noindent [7] Jing, H. and Wang, Y. and Li, Y. Data-Driven Quantum Approximate Optimization Algorithm for Cyber-Physical Power Systems. \textit{arXiv}: 2204.00738 (2022). https://doi.org/10.48550/arXiv.2204.00738.


\bigskip

\noindent [8] Junge, M., Palazuelos, C. On the power of quantum entanglement in multipartite quantum XOR games. \textit{Journal of the London Mathematical Society} \textbf{110}(5) (2024).

\bigskip

\noindent [9] Kubo, K. and Nakagawa, Y.O. and Endo, S. and Nagayama, S. Variational quantum simulations of stochastic differential equations. \textit{Physical Review A} 103: 052425 (2021). https://doi.org/10.1103/PhysRevA.

\noindent 103.052425.


\bigskip



%

\noindent [10] Ostrev, D. Composable, unconditionally secure message authentication without any secret key. IEEE International Symposium on Information Theory \textbf{10}(1109), 622-626 (2019). https://doi.org/10.1109/ISIT.2019.8849510.


\bigskip

\noindent [11] Ostrev, D.. QKD parameter estimation by two-universal hashing. Quantum \textbf{7}, 894 (2023) https://doi.org/10.22331/q-2023-01-13-894.

\bigskip


\noindent [12] Paine, A.E. and Elfving, V.E. and Kyriienko, O. Quantum Kernel Methods for Solving Differential Equations. \textit{Physical Review A} 107: 032428 (2023). https://doi.org/10.1103/PhysRevA.107.032428. 


\bigskip


\noindent [13] Paudel, H.P., Syamlal, M., Crawford, S.E., Lee, Y-L, Shugayev, R.A., Lu, P., Ohodnicki, P.R., Mollot, D., Duan, Y. \textit{Quantum Computing and Simulations for Energy Applications: Review and Perspective. ACS Eng. Au}: 3 151-196 (2022). $\mathrm{https://doi.org/10.1021/ac}$$\mathrm{sengineeringau.1c00033}$.


\bigskip


\noindent [14] Przhiyalkovskiy, Y.V. Quantum process in probability representation of quantum mechanics. \textit{Journal of Physics A: Mathematical and Theoretical} 55: 085301 (2022). https://doi.org/10.1088/1751-8121/ac4b15.


\bigskip

\noindent [15] Perc, M. Statistical physics of human cooperation. \textit{Physics Reports} \textbf{687}: 1-51 (2017). $\mathrm{https://papers.ssrn}$ $\mathrm{.com/sol3/papers.cfm?abstract_id=2972841}$.

\bigskip

\noindent [16] Renner, R., Wolf, S. The Exact Price for Unconditionally Secure Asymmetric Cryptography. In: Cachin, C., Camenisch, J.L. (eds) Advances in Cryptology - EUROCRYPT 2004. EUROCRYPT 2004. Lecture Notes in Computer Science, \textbf{3027}. Springer, Berlin, Heidelberg. $https://doi.org/10.1007/978-3-540-24676-3 7$.

\bigskip

\noindent [17] Rigas, P. Optimal, and approximately optimal, quantum strategies for $\mathrm{XOR^{*}}$ and $\mathrm{FFL}$ games. \textit{arXiv: 2311.12887} (2023), submitted.

\bigskip


\noindent [18] Rigas, P. Quantum strategies, error bounds, optimality, and duality gaps for multiplayer XOR, $\mathrm{XOR}^{*}$, compiled XOR, $\mathrm{XOR}^{*}$, and strong parallel repetiton of XOR, $\mathrm{XOR}^{*}$, and FFL games. 	arXiv:2505.06322, submitted (2025). $\mathrm{
https://doi.org/10.48550/arXiv.2505.06322
}$


\bigskip

\noindent [19] Rigas, P. Error correction, authentication, and false acceptance, probabilities for communication over noisy quantum channels: converse upper bounds on the bit transmission rate. arXiv:2507.03035, submitted (2025). 
https://doi.org/10.48550/arXiv.2507.03035.

\bigskip

\noindent [20] Rigas, P. Parallel repetition of expanded, and multiplayer, Quantum games: anchoring, optimal values, generalized error bounds, dependency-breaking as symmetry-breaking. arXiv: 2508.09380, submitted (2025). 
https://doi.org/10.48550/arXiv.2508.09380.
.

\bigskip

\noindent [21] Rigas, P. Probability distributions over CSS codes: two-universality, QKD hashing, collision bounds, security. arXiv:2510.02402, submitted (2025). https://doi.org/10.48550/arXiv.2510.02402.

\bigskip

\noindent [22] Rigas, P. Composable, unconditional security without a Quantum secret key: public broadcast channels and their conceptualizations, adaptive bit transmission rates, fidelity pruning under wiretaps. arXiv: 2512.19759, submitted (2025). 
https://doi.org/10.48550/arXiv.2512.19759.

\end{document}